\documentclass[12pt]{iopart}

\def\bea{\begin{eqnarray}}
\def\eea{\end{eqnarray}}

\newcommand{\nn}{\nonumber}
\def\beq{\begin{equation}}
\def\eeq{\end{equation}}

\def\pa{\partial}

\newbox\pippobox

\begin{document}

\maketitle

\title{Global monopole solutions in Horava gravity}

\author{Tae Hoon Lee}
\address{Department of Physics and Institute of Natural Sciences, \\
Soongsil University,\\ Seoul 156-743 Korea} \ead{thlee@ssu.ac.kr}

\date{\today}
\begin{abstract}

In Horava's theory of gravity coupled to a global monopole source,
we seek for static, spherically symmetric spacetime solutions for
general values of $\lambda$. We obtain the explicit solutions with
deficit solid angles, in the IR modified Horava gravity model, at
the IR fixed point $\lambda=1$ and at the conformal point
$\lambda=1/3$. For the other values of $1>\lambda>0$ we also find
special solutions to the inhomogenous equation of the gravity model
with detailed balance, and we discuss an possibility of
astrophysical applications of the $\lambda=1/2$ solution that has a
deficit angle for a finite range.

\end{abstract}

\pacs{04.70.Bw, 04.60.Bc, 04.20.Jb}


\maketitle


Since recently Horava proposed a renormalizable gravity theory in
the UV limit \cite{horava}, a lot of related works have been widely
circulated. Studies on Horava-Lifshitz cosmology \cite{jl}, black
hole solutions \cite{bh1, bh2}, and other interesting topics
\cite{topic} have been reported.

In the IR modified Horava theory of gravity \cite{horava} where the
detailed balance condition is softly violated (via the term
proportional to $\omega R$ in Eq. (4) below), we study geometric
structures affected by gravitationally coupled global monopole(GM)
source \cite{vil}. Considering static, spherically symmetric
spacetimes, we obtain solutions to a set of equations derived for
general values of $\lambda$. In this IR modified model, we find the
explicit solutions at the IR fixed point $\lambda=1$ and at the
conformal point $\lambda=1/3$. In both cases deficit solid angles
occur.

For the other values of $1>\lambda>0$ in the case with detailed
balance $\omega=0$, we have new special solutions, in addition to
known general solutions \cite{bh1} to the corresponding homogeneous
equation.
By simple analysis, we show that the GM spacetime in the case
$\lambda=1/2$ can have a deficit solid angle only for a finite range
and that it is asymptotically flat. We discuss an possibility of its
astrophysical applications.


Using the ADM decomposition of the spacetime metric
\beq ds^2=-N^2
dt^2 +g_{ij}(dx^i +N^i dt)(dx^j +N^j dt) \eeq with the lapse $N$ and
shift fields $N^i$,
the IR modified Horava gravity theory is described by the action
\beq S_H= \int dt d^3 x \sqrt{g} N \, [{\cal L}_{K}+{\cal L}_{V}],
\eeq where the kinetic term \beq {\cal L}_{K}=2\kappa^{-2}
(K_{ij}K^{ij}-\lambda K^2) \eeq is made of the extrinsic curvature $
K_{ij}(\equiv {2N}^{-1}(\dot{g}_{ij}-\nabla_i N_j-\nabla_j N_i )), $
its trace $K$ and a parameter $\lambda$.
$ {\cal L}_{V} $ includes all potential terms satisfying the
detailed balance condition \cite{horava}, and it is given by
\bea
{\cal L}_{V} &&=\kappa^2[-\frac{1}{2\zeta^4}C_{ij}C^{ij}+
\frac{\mu}{2\zeta^2} \epsilon^{ijk}R_{il}\nabla_j R_k^l
-\frac{\mu^2}{8} R_{ij}R^{ij} \nn
\\&&
+\frac{\mu^2}{8(1-3\lambda)}(\frac{1-4\lambda}{4} R^2
 + \Lambda R -3 \Lambda^2) -\frac{\mu^2 \omega}{8(1-3\lambda)}R],
\label{origin}\eea
where $\Lambda$ ($<0$) is a cosmological constant and the last term
which violates softly the detailed balance condition is added
\cite{horava}.

When we adopt the static, spherically symmetric metric ansatz as
\beq ds^2 =-N^2(r) \, dt^2 + \frac{dr^2}{f(r)}+r^2(d\theta^2+{\rm
sin}^2 \theta d\phi^2), \eeq we can write $S_H =4\pi \int dt dr
{\cal L}_H$ with the Lagrangian density \bea {\cal L}_H
&&=\frac{N}{q^2 \sqrt{f}}[3\Lambda^2 r^2 +2(\omega-\Lambda)(1-f-r
f')
\\&&+(1-\lambda)\frac{f'^2}{2}
+(1-2\lambda)\frac{(1-f)^2}{r^2}-2\lambda \frac{(1-f)f'}{r} ], \nn
\eea where $q^2=8(3\lambda-1)/(\kappa^2\mu^2)$.

Let us consider a GM source which has the action written upto ${\cal
O}((\pa_j \vec{\Phi})^2)$ \cite{z}
\beq S_{matter}  =-\int dt d^3 x \sqrt{g} N [-\frac{1}{2N^2}
\pa_t\vec{\Phi}\cdot\pa_t\vec{\Phi}+\frac{1}{2} g^{i
j}\pa_i\vec{\Phi}\cdot\pa_j\vec{\Phi}+
\frac{\chi}{4}({\vec{\Phi}}^2-\eta^2)^2 ], \eeq with a dimensionless
coupling constant $\chi$.
We can write $S_{matter}=4\pi \int dt dr \,\,{\cal L}_{matter}$ with
\beq  {\cal L}_{matter}=-\frac{N r^2}{\sqrt{f}}[\frac{1}{2}(f
h'^2+\frac{2h^2}{r^2})+ \frac{\chi}{4}(h^2-\eta^2)^2], \eeq where a
hedgehog ansatz for the GM, $ \vec{\Phi} =h(r) \vec{x}/ r $, is
assumed.

Performing variation of the total action $S_{total}=S_H+S_{matter} $
with respect to $h(r)$, $N(r)$, and $f(r)$ respectively, we obtain
the following equations:


\bea && \frac{\sqrt{f}}{Nr^2} (\frac{Nr^2}{\sqrt{f}} f
h')'=\frac{2h}{r^2} +
\chi(h^2-\eta^2)h, \label{gm} \\
&& (1-\lambda)\frac{f'^2}{2}
 +(1-2\lambda)\frac{(1-f)^2}{r^2}  -2\lambda \frac{(1-f)f'}{r}
 + 2(\omega-\Lambda)(1-f-r f')
 +3\Lambda^2 r^2 \nn \\&&=q^2 r^2
[\frac{f h'^2}{2}+\frac{h^2}{r^2}+\frac{\chi}{4}(h^2-\eta^2)^2] ,
\label{N}\\
&& (\frac{N}{\sqrt{f}} )' \, [-2\lambda
\frac{(1-f)}{r}+(1-\lambda)f' -2r(\omega-\Lambda)]  \label{f} \\&&=
-\frac{N}{\sqrt{f}} [2(1-\lambda)\frac{(1-f)}{r^2}+(1-\lambda)f''
+q^2 \frac{ r^2 h'^2}{2}].\nn \eea


With the solution to Eq. (\ref{gm})
 \beq h(r) = \eta \label{asy}\eeq valid for the outside of the GM core, $r >
\chi^{-1/2} \eta^{-1} $ \cite{vil}, the solutions to the other
equations (\ref{N}) and (\ref{f}) for various values of $\lambda$
are given as follow.


\section{$\lambda=1$ case}
In this $\lambda=1$ case where Horava's theory coincides with
Einstein's general theory of relativity in IR limit, Eq. (\ref{asy})
gives us the simple solution to Eq. (\ref{f}) as $
{N}/{\sqrt{f}}=1$.
Putting $1-f \,  \equiv -(\omega-\Lambda)r^2+X^{1/2}$, we can
rewrite the remaining equation (\ref{N})
 as
\beq
3\omega(\omega-2\Lambda)r^2+q^2\eta^2=\frac{X'}{r}-\frac{X}{r^2},
\eeq whose solution is \beq
f=1+(\omega-\Lambda)r^2-\sqrt{\omega(\omega-2\Lambda)r^4+q^2\eta^2
r^2+\beta r}. \label{gms} \eeq
Note that Eq. (\ref{gms}) would be the same as the result of Ref.
\cite{MP} if there were not the additional new term $q^2 \eta^2
r^2$. In the limit $r>>\sqrt{q^2\eta^2/[\omega(\omega-2\Lambda)]}$
and $r>>[\beta/\{\omega(\omega-2\Lambda)\}]^{1/3} $, Eq. (\ref{gms})
can be approximated as \beq
f=1-\frac{q^2\eta^2}{2\sqrt{\omega(\omega-2\Lambda)}}+\frac{\Lambda_{eff}}{2}
\, r^2 -\frac{\beta}{2\sqrt{\omega(\omega-2\Lambda)} \, r}, \eeq
which can be compared with the Schwarzschild-AdS black hole carrying
a GM charge $q\eta$ and a mass $M\simeq
\beta/[4\sqrt{\omega(\omega-2\Lambda)}]$. Here a effective
cosmological constant $\Lambda_{eff}\equiv
2[(\omega-\Lambda)-\sqrt{\omega(\omega-2\Lambda)}](\simeq\Lambda^2/\omega$
for $-\Lambda<\omega$) and a deficit angle
$q^2\eta^2/[2\sqrt{\omega(\omega-2\Lambda)}]=8\eta^2/[\kappa^2\mu^2
\sqrt{\omega(\omega-2\Lambda)}]$.

\section{$1>\lambda$ case} With Eq. (\ref{asy}) in the case where
$1>\lambda$, we can rewrite Eq. (\ref{N}) as \beq
\frac{1-\lambda}{2}(\frac{d
Y}{du})^2+\frac{1-3\lambda}{1-\lambda}Y^2=3\omega(\omega-2\Lambda)e^{(\frac{4\lambda}{1-\lambda}+4)u}
  +q^2\eta^2e^{(\frac{4\lambda}{1-\lambda}+2)u} \label{gen} \eeq
with $u={\rm ln}\, r$ and
$Y(u(r))=r^{2\lambda/(1-\lambda)}(1-f(r))+(\omega-\Lambda)r^{2/(1-\lambda)}$

\subsection{$\lambda=\frac{1}{3}$ case}
In the case $\lambda=1/3$ where it is allowed for us to get a
nontrivial conformal limit \cite{conf}, the set of equations in Eqs.
(9)-(\ref{f}) obtained from Eqs. (4) and (8) can be replaced by the
same form with only substitution $q^2\rightarrow
q^2/(3\lambda-1)=8/(\kappa^2 \mu^2)$,
and we have their solutions \beq
f=1+(\omega-\Lambda)r^2-\frac{2M}{r} -
\frac{\sqrt{3}}{9\omega(\omega-2\Lambda)\,r}[
\frac{8\eta^2}{\kappa^2\mu^2} +3\omega(\omega-2\Lambda)
r^2]^{\frac{3}{2}}, \label{conf} \eeq and $N^2=r^2 f(r).$  Eq.
(\ref{conf}) goes to \beq
f=1-\frac{4\eta^2}{\kappa^2\mu^2\sqrt{\omega(\omega-2\Lambda)}}
+\frac{\Lambda_{eff}}{2}\, r^2-\frac{2M}{r}, \eeq in the region
$r>>\sqrt{8}\eta/[\kappa\mu\sqrt{\omega(\omega-2\Lambda)}] .$ In
this large $r$ limit, $f(r)$ (of this case $\lambda=1/3$) is almost
the same as one of the $\lambda=1$ case (in Eq. (15)) except
different values of the deficit angle, while the lapse functions
$N(r)$ in these cases are very different from each other.

When there is no GM source, we have \beq f=1+
\frac{\Lambda_{eff}}{2} \, r^2-\frac{2M}{r}, \eeq and $N^2=f(r).$

\subsection{$1 > \lambda > \frac{1}{3}$ case}

From now on, we consider the case with detailed balance condition
({\it i.e.} $\omega=0$) and without $q^2$ rescaling which is done in
the case $\lambda=1/3$. Eq. (\ref{gen}) can be written as a simple
inhomogeneous equation \beq (\frac{d Y}{d U})^2=AY^2+Be^U,
\label{yeq} \eeq where $ U \equiv \gamma u $,
$A=2(3\lambda-1)/(\gamma^2 (1-\lambda)^2) $, $B=2q^2
\eta^2/(\gamma^2 (1-\lambda))$, and
$\gamma=2(1+\lambda)/(1-\lambda)$.

The solution is
\beq f=1 -\Lambda r^2 -q\eta\sqrt{\frac{1- \lambda}{3\lambda-1}}
\frac{r}{R(r)} ,\label{tan} \eeq where with a constant $c_R$ \bea
{\rm ln} \, r(R) \,(=\frac{U}{\gamma})=&& c_R
-\frac{\sqrt{2(3\lambda-1)}}{\lambda-3} {\rm ln} \frac{\sqrt{1+R^2}
-1}{R}\nn\\&&+\frac{1+\lambda}{\lambda-3}{\rm ln} \frac{\vert 1-
\frac{\sqrt{2(3\lambda-1)(1+R^2)}}{1+\lambda}\vert}{R}.\eea
From the last equation we can estimate the asymptotic behavior of
Eq. (\ref{tan}) as; $r/R\simeq 0$ for large $R$, while, for small
$R$, $r/R\propto r^{1-1/n(\lambda)}$ with $n(\lambda)\equiv
-1+(4+\sqrt{2(3\lambda-1)})/(3-\lambda)$ and
$-1<1-1/n(\lambda)<1/2$. Since especially $r/R = constant$ when
$n(\lambda=1/2)=1$, we may have a deficit angle for a finite range
$r<r_0$ (with a constant $r_0$) in the case $\lambda=1/2$.

The lapse function $N=\sqrt{f(r)} M$ with \bea &&ln
\frac{M}{\sqrt{2(1-\lambda)q^2\eta^2 +2(3\lambda-1)r^{-2}(1-f
-\Lambda r^2)^2 }}\nn\\&&=\int dr [ \frac{2\lambda}{(1-\lambda) r}
\label{int}
\\&&-\frac{2(3\lambda-1)(1-f-\Lambda r^2)}{(1-\lambda) r^2 \sqrt{2(1-\lambda)q^2\eta^2
+2(3\lambda-1)(1-f -\Lambda r^2)^2/r^{2} } }]. \nn  \eea


When $q\eta=0$, the lapse \cite{bh1}  \beq N=\sqrt{f(r)} \,
r^{\frac{1+3\lambda \pm 2 \sqrt{2(3\lambda-1)}}{1-\lambda}} \eeq is
obtained from Eq. (\ref{int}) with the substitution of the term
$1-f-\Lambda r^2$ by
$r^{\frac{-2\lambda\pm\sqrt{2(3\lambda-1)}}{1-\lambda}}$ \cite{bh1}
which is a solution to the corresponding homogeneous equation of
(20) (instead of the last term in Eq. (21)).

\subsection{$\lambda < \frac{1}{3}$ case} When $\lambda <
\frac{1}{3}$, Eq. (\ref{yeq}) is replaced by  \beq (\frac{d Y}{d
U})^2=-\alpha  Y^2+Be^U, \eeq with $\alpha=2(1-3\lambda)/(\gamma^2
(1-\lambda)^2)>0$ and $B>0$ given below Eq. (20). This inhomogeneous
equation has a (special) solution
 \beq f=1 -\Lambda r^2
-q\eta\sqrt{\frac{1- \lambda}{1-3\lambda}}  r \, I(r) ,\eeq where
\bea  {\rm ln} \,
r(I)&&=\frac{U}{\gamma}=c_I-\frac{\sqrt{2(1-3\lambda)}}{\lambda-3}
{\rm arctan}\frac
{I}{\sqrt{1-I^2}}\nn\\&&+\frac{1+\lambda}{\lambda-3}{\rm ln} \vert
\frac{\sqrt{2(1-3\lambda)(1-I^2)}}{1+\lambda} -I \vert. \eea The
lapse function in this case can be obtained by the similar method as
we have done in Eq. (23).


In summary, we have studied the IR modified Horava theory of
gravity. In static, spherically symmetric spacetimes, we obtain
exact solutions (valid outside a GM core) for general values of
$\lambda$ to the equations of gravity coupled to the GM.
As we can see from Eqs. (15) and (18) obtained in the large $r$
limit, in the cases $\lambda=1$ and $\lambda=1/3 $ we have deficit
angles as in Einstein's theory of gravity coupled to the GM
\cite{vil}. $f(r)$ of the case $\lambda=1/3$ (in Eq. (18)) is almost
the same as one of the $\lambda=1$ case (in Eq. (15)) except
different values of the deficit angle. We also have the explicit
solutions of the lapse function in both cases.

In the case $1>\lambda>1/3$ we have studied the Horava model with
detailed balance and obtained special solutions, in addition to
known general solutions \cite{bh1} to its homogeneous equation.
When especially $\lambda=1/2$, $r/R\simeq constant$ for $r<r_0$ as
seen in Eqs. (21), (22) and below, and we can have a GM spacetime
that has a deficit solid angle for a finite range and is
asymptotically flat, which is different from the GM spacetime in
Einstein's theory of gravity \cite{vil}. This might be more helpful
for us, with the GM as Refs. \cite{nuc, thl}, to explain near
flatness of rotation curves of galaxies, which appears over a finite
range $0 << r < r_0$.

To explain the near flatness of rotation curves in preceding models
using GM with an energy density proportional to $ r^{-2} $, we need
nonlinear coupling between gravity and the GM as nonminimal coupling
in Ref. \cite{nuc} or Brans-Dicke field coupling. In the latter
case, as discussed below Eq. (4) of Ref. \cite{reply}, it can be
yielded by the finite range, logarithmic gravitational potential
that is derived from the Brans-Dicke field equation.
For the rotation velocity formula to be valid only for the finite
range given by the galactic halo radius $r_0$, the responsible GM
field should vanish at distance larger than $r_0$ due to
interactions with the nearest topological defect such as
anti-monopole, in the way that the GM field lines can be absorbed
into the anti-monopole core, as argued in Ref. \cite{reply}.

Instead, if we study further (possibly considering Brans-Dicke field
coupling to Horava gravity \cite{bdhlghs}) the $\lambda=1/2$ Horava
gravity solution given below Eq. (22) having a finite range deficit
angle, more natural explanations for the near flatness can be
possible. This kind of $r$-dependent, deficit solid angle was
obtained in Brans-Dicke gravity theory \cite{miracle}, by studying
the quantum effects \cite{his} due to the GM, which can be expressed
as quadratic in curvature as if the Horava gravity with detailed
balance.
%
When we almost complete our study, we see Ref. \cite{skku} that has
reported results including some information consistent with ours in
the section 1.
We have not considered higher derivative terms of GM fields in Eq.
(7) for simplicity. Even if we add these terms
$(\pa_j(\pa_k\pa^k)^{(z-1)/2} \vec{\Phi})^2 $ ($1< z\leq 3$)
\cite{z, skku}, with the vacuum solution Eq. (12) our main results
are not changed in the leading $1/r$ approximation.

\section*{Acknowledgments}

We thank Professor C. Lee, colleagues P. Oh and J. Lee, J. G. Lee,
and others for helpful discussions. This work was supported by the
Soongsil University Research Fund.


\begin{thebibliography}{}

\bibitem{horava}
Horava P 2009 {\it Phys. Rev.} D {\bf 79} 084008, arXiv:0901.3775
[hep-th]
%
\bibitem{jl}
Calcagni G 2009 {\it  JHEP} {\bf 0909} 112, arXiv:0904.0829
[hep-th]\\
%
Kiritsis E and Kofinas G 2009 {\it Nucl. Phys.} B {\bf 821} 467,
arXiv:0904.1334 [hep-th]\\
%
Brandenberger R 2009 {\it Phys. Rev.} D {\bf 80} 043516,
arXiv:0904.2835 [hep-th]
%
\bibitem{bh1}
Lu H, Mei J and Pope C N 2009 {\it Phys. Rev. Lett.} {\bf 103}
091301, arXiv:0904.1595 [hep-th]
%
\bibitem{bh2}
Cai R-G, Cao L-M and Ohta N 2009 {\it Phys. Rev.} D {\bf 80} 024003,
arXiv:0904.3670 [hep-th]
%
\bibitem{topic}
Dutta S and Saridakis E N 2010 {\it JCAP} {\bf 1005} 013, arXiv:1002.3373 [hep-th]\\
Takahashi T and Soda J 2009 {\it Phys. Rev. Lett.} {\bf 102} 231301,
arXiv:0904.0554 [hep-th]\\
%
Mukohyama S 2009 {\it JCAP} {\bf 0906} 001, arXiv:0904.2190
[hep-th]\\
%
Kehagias A and Sfetsos K 2009 {\it Phys. Lett.} B {\bf 678} 123,
arXiv:0905.0477 [hep-th]\\
%
Charmousis C, Niz G, Padilla A and Saffin P A 2009 {\it JHEP} {\bf
0908} 070, arXiv:0905.2579 [hep-th]
%
\bibitem{z}
Tang J-Z 2009 arXiv:0911.3849 [hep-th]\\
Chen B and Huang Q-G 2010 {\it Phys. Lett.} B {\bf 683} 108, arXiv:
0904.4565 [hep-th]\\
Horava P 2008 arXiv:0811.2217 [hep-th]
%
\bibitem{vil}
Barriola M and Vilenkin A 1989 {\it Phys. Rev. Lett.} {\bf 63} 341
%
\bibitem{MP}
Park M-I 2009 {\it JHEP} {\bf 0909} 123, arXiv:0905.4480 [hep-th]
%
\bibitem{conf}
Capasso D and Polychronakos A P 2010 {\it Phys. Rev.} D {\bf 81}
084009, arXiv:0911.1535 [hep-th]
%
\bibitem{nuc}
Nucamendi U, Salgado M and Sudarsky D 2000 {\it Phy. Rev. Lett.}
{\bf 84} 3037
%
\bibitem{thl}
Lee T H and Lee B J 2004 {\it Phys. Rev.} D {\bf 69} 127502
%
\bibitem{reply}
Lee T H and Lee B J 2006 {\it Phys. Rev.} D {\bf 73} 128502,
gr-qc/0604046
%
\bibitem{bdhlghs}
Lee J, Lee T H and Oh P, arXiv:1003.2840 [hep-th]
%
\bibitem{miracle}
Rahaman F and Ghosh P 2008 {\it Mod. Phys. Lett.} A {\bf 23} 2763,
arXiv:0801.2622 [gr-qc]
%
\bibitem{his}
Hiscock W A 1990 {\it Class. Quant. Grav.} {\bf 7} 6235
%
\bibitem{skku}
Kim T and Lee C O, arXiv:1002.0784 [hep-th]\\
Kim S-S, Kim T and Kim Y 2009 {\it Phys. Rev.} D {\bf 80} 124002,
arXiv:0907.3093 [hep-th]
%
\end{thebibliography}
\end{document}